\documentclass[11pt]{article}
\usepackage{latexsym}

\usepackage{amssymb}
\usepackage{amsmath}

 \usepackage{epsfig}
 \usepackage{graphicx}

\newcommand{\be}{\begin{equation}}
\newcommand{\ee}{\end{equation}}
\newcommand{\ba}{\begin{eqnarray}}
\newcommand{\ea}{\end{eqnarray}}
\newcommand{\baa}{\begin{array}}
\newcommand{\eaa}{\end{array}}
\newcommand{\nn}{\nonumber \\}
\newcommand{\nr}[1]{(\ref{#1})}
\newcommand{\la}[1]{\label{#1}}

\newcommand{\rmi}[1]{{\mbox{\scriptsize #1}}}
\newcommand{\fr}[2]{{\frac{#1}{#2}\,}}
\newcommand{\fra}[2]{\textstyle{\frac{#1}{#2}\,}}  
\newcommand{\mn}{{\mu\nu}}
\newcommand{\bfx}{{\bf x}}

\newcommand{\tinymsbar}{{\overline{\mbox{\tiny\rm{MS}}}}}

\def\CL{{\cal L}}

\def\gsim{\raise0.3ex\hbox{$>$\kern-0.75em\raise-1.1ex\hbox{$\sim$}}}
\def\lsim{\raise0.3ex\hbox{$<$\kern-0.75em\raise-1.1ex\hbox{$\sim$}}}

\begin{document}

\begin{titlepage}
\begin{flushright}
HIP-2009-25/TH\\
November 2009\\
\end{flushright}
\begin{centering}
\vfill

{\Large{\bf A gauge/gravity duality model for gauge theory thermodynamics}}

\vspace{0.8cm}

\renewcommand{\thefootnote}{\fnsymbol{footnote}}

J. Alanen$^{\rm a,b}$\footnote{janne.alanen@helsinki.fi},
K. Kajantie$^{\rm a,b}$\footnote{keijo.kajantie@helsinki.fi},
V. Suur-Uski$^{\rm a,b}$\footnote{ville.suur-uski@helsinki.fi}

\setcounter{footnote}{0}

\vspace{0.8cm}

{\em $^{\rm a}$%
Department of Physics, P.O.Box 64, FI-00014 University of Helsinki,
Finland\\}
{\em $^{\rm b}$%
Helsinki Institute of Physics, P.O.Box 64, FI-00014 University of
Helsinki, Finland\\}

\vspace*{0.8cm}

\end{centering}

\noindent
We study a gauge/gravity model for the thermodynamics of a 
gauge theory with one running coupling.
The gravity side contains an ansatz for the metric and a scalar field, on the
field theory side one starts by giving an ansatz for the beta function
describing the scale dependence of the coupling. 
The model is based on relating the scale to the extra dimensional coordinate 
and the beta function to the gravity fields, thereby
also determining the scalar field potential. We study three different forms
of beta functions of increasing complexity and give semianalytic solutions
describing first order or continuous transitions. 

\vfill \noindent

%

\vspace*{1cm}

\noindent


\vfill

\end{titlepage}

\section{Introduction}
The prototype gauge/gravity duality relates finite temperature
conformally invariant ${\cal N}=4$ supersymmetric
Yang-Mills theory to 5-dimensional AdS space (times S$_5$) with a black hole \cite{gkp,gkt}.
To extend this duality to usual SU($N_c$) Yang-Mills theory at finite $T$, on which there is
ample lattice Monte Carlo \cite{boyd, teper1, teper2,panero} and
even (with quarks included) experimental data,
one of the first tasks is to see how conformal invariance can be broken, how new scales can
be introduced. A large number of models have been suggested for this \cite{herzog,kty,bb,
kiri1,kiri2,kiri3,kiri4,gubsernellore,gubsernellorepufu,dewolferosen,hohlerstephanov}.
This article is one more, very closely related to \cite{kiri3}.

The basic building blocks of the gravity dual discussed here are ($z$ is the 5th coordinate, the 
boundary is at $z=0$)
the conformal factor of the 5d metric $b(z)$ (=$\CL/z$, $\CL$ = AdS radius, in the conformal case),
the black hole factor $f(z)$, $f(z_h)=0$ ($=1-z^4/z_h^4$ in the conformal case), the scalar
field $\phi(z)=\log\lambda(z)$ (= 0 in the conformal case) and the scalar field potential
$V(\phi)$ in the gravity action (= $12/\CL^2$ in the conformal case).
Assume now the boundary theory one is
searching gravity dual for has a running coupling $g^2(\mu)$ with a known beta function:
$\mu\, dg^2(\mu)/d\mu=\beta(g^2)$. The key idea now is to firstly identify 
$\mu$ (in some units) with $b(z)$ so that large
energy scales, ultraviolet, UV, correspond to 
$z\to0,\,b(z)\sim1/z\to\infty$ and small scales, infrared, IR, to large
$z$. In the IR $b$ may approach a constant or vanish. 
Secondly, one identifies $\lambda=e^\phi$ with $g^2$. One thus has, in addition to three
Einstein equations, one more equation $b\,d\lambda/db=\beta(\lambda)$
relating the metric and the scalar.
In the analysis of \cite{kiri1}-\cite{kiri4}
the scalar potential $V(\phi)$ in the gravity action is so constructed that
a desired beta function is obtained in the UV and confinement in the IR.

In this article, we take an extreme view and use the two Einstein equations not containing
$V$ and the equation $b\,d\lambda/db=\beta(\lambda)$ to completely solve the problem
and simply regard the Einstein equation
containing $V(\phi)$ as an equation determining $V(\phi)$. We shall mainly
concentrate on studying 
a simple powerlike beta function ansatz, $\beta(\lambda)=-\beta_0\lambda^q$.
Here $q$ is a parameter, $q\ge1$ corresponds to a theory confining in the IR,  
$q=2$ would correspond to one-loop Yang-Mills running and $q=10/3$ has
in \cite{aks} been shown to reproduce the main
features of the detailed analysis in \cite{kiri4}. 
Further, we shall modify this ansatz so that its 
large $\lambda$ behavior is constrained to be $-\fra32\lambda(1+\alpha/\log\lambda+..)$.
First order transitions are obtained for $q>1$ and $\alpha>0$, 
$\alpha=0$ describes a continuous transition.

The theoretical basis of this model has been laid in \cite{kiri1}-\cite{kiri4}, starting
from a potential $V(\phi)$. The chief virtue of the approach in this article,
starting from $\beta(\lambda)$, is that
semianalytic formulas, requiring only numerical evaluation of integrals, are obtained.
Details of SU($N_c$) gauge theory thermodynamics can be described with 
$q,\alpha$ and two more parameters:
a dimensionless $\CL^3/G_5$ fixed by $p(T)/T^4$ at large $T$ and
an energy scale $\Lambda$ fixed by $T_c$, the transition temperature between
high and low $T$ phases. Note that, in particular, 
the numerical value of the parameter $\beta_0$
in the beta function ansatz, never enters for physical quantities.

\section{The model}
One starts from the gravity + scalar action (in the Einstein frame and in standard notation)
\be
S={1\over16\pi G_5}\left\{\int d^5x\,\sqrt{-g}\left[R-\fra43(\partial_\mu\phi)^2+V(\phi)\right]
-2\int d^4x\,\sqrt{-\gamma}K\right\}.
\la{Eframeaction}
\ee
By writing
\be
g^s_{\mn}=e^{\fr43\phi}g^E_{\mn},
\la{stringeinstein}
\ee
it can also be written in the string frame.
One now assumes a metric ansatz
\be
ds^2=b^2(z)\left[-f(z)dt^2+d\bfx^2+{dz^2\over f(z)}\right].
\la{ansatz}
\ee
The four functions $b(z), f(z)$ in the metric, the scalar field $\phi(z)$ and the
potential $V(\phi(z))$ are then
determined as the solutions of the three field equations following from \nr{Eframeaction}:
\ba
&&6{\dot b^2\over b^2}+3{\ddot b\over b}+3{\dot b\over b}{\dot f\over f}={b^2\over f}V(\phi),\label{eq1}\\
&& 6{\dot b^2\over b^2}-3{\ddot b\over b}={\fra43} \dot\phi^2,\label{eq2}\\
&&{\ddot f\over \dot f}+3{\dot b\over b}=0,\label{eq3}
\ea
($\dot b\equiv b'(z)$, etc.) and from a fourth equation,
\be
\beta(\lambda)=b{d\lambda\over db},\quad \lambda(z)  = e^{\phi(z)}\sim  g^2N_c,
\label{crucial}
\ee
where $\beta(\lambda)$ is the beta function of the field theory one is seeking the
gravity dual for. To begin with, we use the very simple beta function
\be
\beta(\lambda)=-\beta_0\lambda^q
\label{betaappro}
\ee
and solve the equations in the order \nr{crucial}, \nr{eq2} and \nr{eq3}. After this,
Eq. \nr{eq1} gives the potential $V(\phi)$. In Sections \ref{conttrans} and \ref{1storder}
slightly modified beta functions are studied.

In \cite{kiri3,kiri4} the starting point was the potential
\ba
V(\phi)&=&{12\over\CL^2}\left\{1+V_0\lambda+V_1\lambda^{4/3}[\log(1+V_3\lambda^2)]^{1/2}\right\}
\label{pot}\\
&\approx&{12\over\CL^2}\left[1+V_0\lambda+V_1\sqrt{V_3}\lambda^{7/3}+{\cal O}(\lambda^{13/3})\right],
\label{potappro}
\ea
where the parameter values used were
\be
V_0=0.04128,\quad V_1=14.3,\quad V_3=170.4,\quad \lambda(z=1)=0.0242254. \label{params}
\ee
$V_0$ matters only in far UV and can be set to zero for thermodynamics, which is dominated
by the $\lambda^{7/3}$ term \cite{aks}. The equations of motion \nr{eq1}-\nr{eq3} were
solved numerically in \cite{kiri4} and \nr{betaappro} can then
be used to determine the beta function \cite{aks}.
Here we proceed in the opposite direction, starting from the beta function. The prototype
values of $q$ would be $q=2$ corresponding to the $V_0$ term in \nr{potappro} and
$q=10/3$ corresponding to the thermodynamically dominant $\lambda^{7/3}$ term
(see Eq. \nr{Vlim} below). In general, $q>1$ and the limit $q=1$ plays a special role,
e.g., the latent heat vanishes in this limit.

\subsection{The fourth equation.}
First, from Eq.\nr{crucial} and for $q>1$
\be
\log{b\over b_0}={1\over(q-1)\beta_0\lambda^{q-1}}\equiv Q,
\la{betalam}
\ee
where $b_0$ is a constant, the analogue of $\Lambda_\rmi{QCD}$, the scale at which
$\lambda$ diverges. It will prove to be convenient to use Q as the extra dimensional
variable and relate $b,\lambda,f$ and $z$ to it. The numerical value of $\beta_0$
and the normalisation of $\lambda$ do not matter, only the combination in \nr{betalam}
enters. We shall also abbreviate
\be
a\equiv{16\over9(q-1)^2},\qquad A\equiv\fr{1}{2}\sqrt{a}= {2\over3(q-1)}.
\ee

An important characteristic of \nr{betalam} is that in the IR, 
$\lambda\to\infty,\,\,Q\to0$, $b$
approaches a non-zero constant $b_0>0$. We shall in Section \ref{conttrans} study the
case when $b$ vanishes proportionally to a power of $Q$ and in Section \ref{1storder} the case
of $b$ vanishing $\sim$ powers of $Q$ and $\log Q$. Then the simple relation \nr{betalam}
between $b$ and $Q$ is replaced by the more complicated one in \nr{betalamfull}.

\subsection{The second equation.}
To integrate the second equation one introduces
\be
W= -\dot b/b^2,
\la{defW}
\ee
which simplifies \nr{eq2} to the form
$$
b\dot W=\fra49\dot\phi^2=\fra49\dot\lambda^2/\lambda^2.
$$
Writing $\dot W=dW/d\lambda\cdot\dot\lambda$ cancels one power of $\dot\lambda$
and replacing the remaining $\dot\lambda$ by the beta function using the basic
equation \nr{crucial} just produces to the right hand side the definition
\nr{defW} of $W$. The equation then integrates to give
\be
W(\lambda)=W(0)\exp\left(-\fra49\int_0^\lambda d\bar\lambda{\beta(\bar\lambda)
\over\bar\lambda^2}\right)={1\over\CL}\exp{4\over9(q-1)^2\log(b/b_0)}\equiv
{1\over\CL}\exp{a\over4Q}.
\label{W}
\ee
The normalisation $W(0)=1/\CL$ follows from the requirement that the boundary
be asymptotically AdS, $V(0)=12/\CL^2$ (see Eq. \nr{V} below).
Now that $W$ is known, the second integration leading to $b=b(z)$ is performed
by writing \nr{defW} in the form
\be
dz={db\over -b^2W}={dQ\over -bW}=-{\CL\over b_0}dQ\exp\left[-Q-{a\over4Q}\right].
\la{dzdQ}
\ee
or, if one wants $\lambda=\lambda(z)$,
\be
{d\lambda\over dz}={db\over dz}{\beta\over b}=-\beta(\lambda)b(\lambda)W(\lambda).
\la{zlab}
\ee
Here the important energy scale
\be
\Lambda={b_0\over\CL}
\ee
has appeared and one has
\be
\Lambda z=I(Q,\fra14 a),
\la{zQ}
\ee
where
\ba
I(Q,a)&=&\int_Q^\infty dy\exp\left[-y-{a\over y}\right]\\
&=&\sum_0^\infty {1\over n!}\left(-a\right)^n \Gamma(1-n,Q)\\
&=&\exp\left[-Q-{a\over Q}\right]\left(1+{a\over Q^2}-{2a\over Q^3}+
{a^2+6a\over Q^4}- {6a^2+24a\over Q^5}+..\right),\\
I(0,a)&=&2\sqrt{a}K_1(2\sqrt{a}),\la{i0}
\label{gammaexp}
\ea
A few terms of the asymptotic expansion at large $Q$ are exhibited here.
The behavior of the integral $I(Q,a)$ is as shown in Fig. \ref{fig:izT}.
The $Q=0$ limit together with \nr{i0} implies that
\be
z\le{1\over\Lambda}\sqrt{a}K_1(\sqrt{a})={1\over\Lambda}\,{4\over9(q-1)}
K_1\left({4\over9(q-1)}\right);
\la{endz}
\ee
a "wall" has been dynamically generated via the assumption \nr{crucial}.

\begin{figure}[!tb]
\begin{center}

\includegraphics[width=0.5\textwidth]{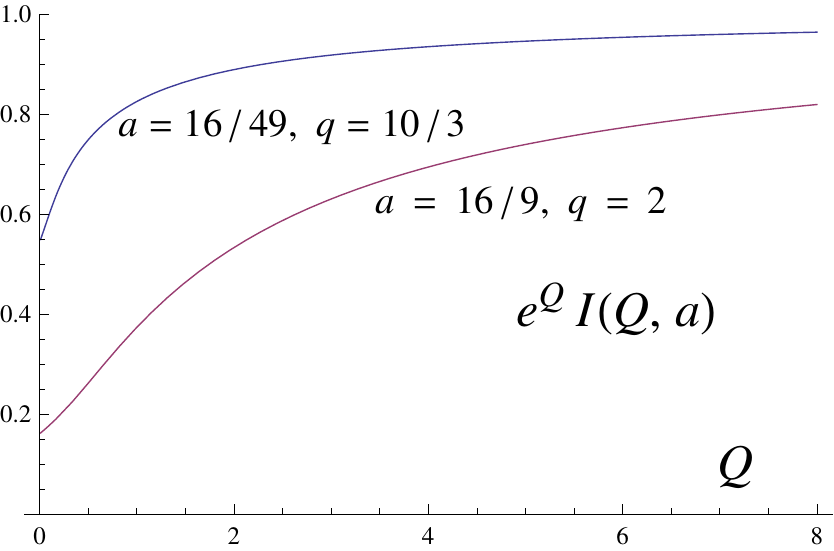}

\end{center}

\caption{\small Plot of the integral $e^QI(Q,a)$ for two values of $a$.
\la{fig:izT}
}
\end{figure}

\subsection{The third equation.}
The third equation (together with the first one) always has the trivial
solution $\dot f=0,\,\,f=1$. This will be the low temperature phase with
zero free energy. It is confining for $q>1$: the confinement criterion
is that in the string frame the metric factor $b_s=b\lambda^{2/3}$ have a
minimum at some $z$ (see Fig.~\ref{fields_z}).
Using \nr{betaappro} this converts to the condition that
the equation $\beta(\lambda)+\fra32\lambda=0$ have a solution. This is so
for $q>1$ and one finds \cite{aks}, e.g., the string tension
\be
\sigma={\CL^2\Lambda^2\over2\pi\alpha'}\left({3e\over 2\beta_0}\right)^{4/(3q-3)}.
\ee

To have a nontrivial solution, one
changes $dz$ to $dQ$ using \nr{dzdQ} and directly integrates
the third equation \nr{eq3} to give
\ba
f(z)&=&C_1+C_2\int_0^z{d\bar z\over b^3(\bar z)}=C_1+C_2{\CL\over b_0^4}I(4Q,a)
\nonumber \\
&=& 1-\int_0^z{d\bar z\over b^3(\bar z)}/\int_0^{z_h}{d\bar z\over b^3(\bar z)}
=1-{I(4Q,a)\over I(4Q_h,a)},
\la{fz}
\ea
where $C_2$ is fixed by introducing a scale $Q_h=Q(z_h)$ at which $f$ vanishes, $f(Q_h)=0$,
and $C_1=1$ by the fact that $b(z)=\CL/z,\,z\to0$ demands $f(z=0)=f(Q=\infty)=1$.

\begin{figure}[!tb]
\begin{center}

\includegraphics[width=0.5\textwidth]{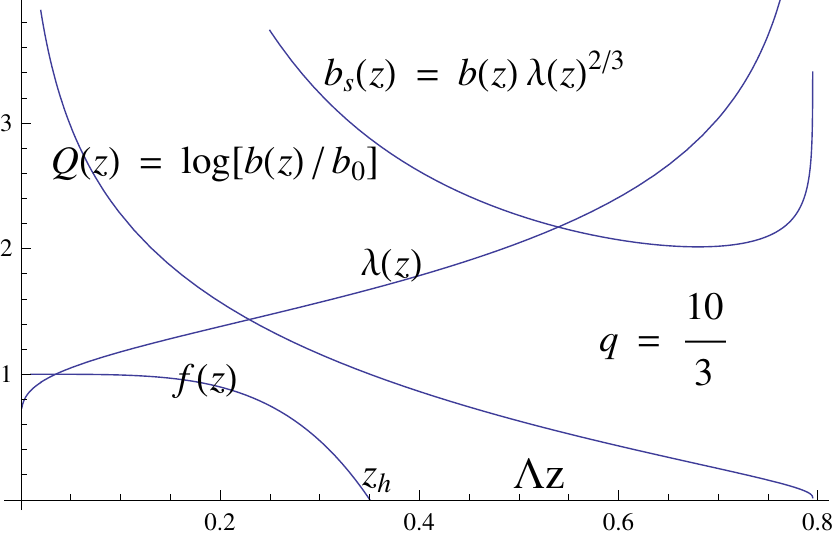}

\end{center}

\caption{\small Field configurations for $q=10/3$ 
as functions of the extra dimensional coordinate. First
$Q=Q(z)$ is computed from \nr{zQ}, then $b(z)$ is obtained from \nr{betalam},
$z$ extends up to \nr{endz}. 
Then $\lambda=\lambda(z)=1/[(q-1)\beta_0Q(z)]^{1/(q-1)}$ is 
obtained from $Q(z)$ by inverting \nr{betalam} and setting $(q-1)\beta_0=0.3$. 
These give the string frame $b_s(z)$. 
Finally $f(z)$ follows from \nr{fz} setting 
$Q_h=1$. The choices of parameters have no effect on the thermodynamics. 
\la{fields_z}
}
\end{figure}

\subsection{The first equation, the potential $V(\phi)$.}
Now that we have all the functions solved, we can insert them to the
first equation \nr{eq1} and obtain the potential
\ba
V(\lambda)&=&12fW^2\left[1-\left({\beta\over3\lambda}\right)^2\right]-3{\dot f\over b}W
\label{V}\\
&=&{12\over\CL^2}\exp\left({a\over4Q}\right)
\left[\left(1-{I(4Q,a)\over I(4Q_h,a)}\right)\exp\left({a\over4Q}\right)
\left(1-{a\over16Q^2}\right)+{\exp(-4Q)\over I(4Q_h,a)}\right]
\la{Vexpl}\\
&=&{12\over\CL^2}\left(1+{a\over8Q}+{2a^2-a\over16Q^2}+..\right) =
{12\over\CL^2}\left(1+{8\beta_0\over 9(q-1)}\lambda^{q-1}+..\right).
\la{Vlim}
\ea
Eq. \nr{Vexpl} is plotted in Fig. \ref{fig:V} for $a=16/(9(q-1)^2)=16/49$
by choosing some values for $Q_h$. $V$ is always positive in the form \nr{Vexpl}:
there is a range beyond the horizon where $f(1-(\beta/(3\lambda))^2)<0$ but the last
term cancels this negative part.
One sees how the potential agrees with
the potential in \cite{kiri3,kiri4} for $Q>Q_h$, $z<z_h$. Analytically,
the $Q\gg Q_h$, $\lambda\ll1$, $z\ll z_h$ limit \nr{Vlim} coincides with the same limit of
the potential in \cite{kiri3,kiri4}. Beyond the horizon
$V$ starts growing rapidly, in fact, for $Q\ll Q_h$, $V(Q)\sim Q^{-2}\exp[a/(2Q)]\sim
\lambda^{2q-2}\exp[\fra8{9(q-1)}\beta_0\lambda^{q-1}]$.

The solution $b(z),\,\phi(z),\,f(z)$ for the thermal case is given concretely by the
equations above. To get the vacuum solutions $b_0(z),\,\phi_0(z),\,f=1$ one should
now solve Eqs. \nr{eq1} and \nr{eq2} with $f=1$ and using the potential \nr{Vexpl}.
We do not need these solutions, only that they exist.

\begin{figure}[!tb]
\begin{center}

\includegraphics[width=0.5\textwidth]{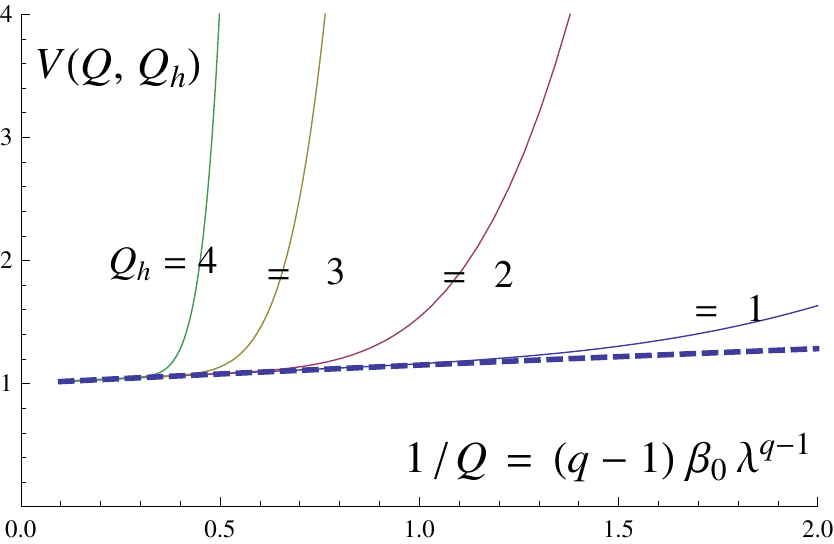}

\end{center}

\caption{\small $V(Q)$ from \nr{Vexpl} for $q=10/3$.
The dashed line is the full potential from \cite{kiri4},
indistinguishable from the $Q>Q_h$ limit \nr{Vlim}.
\la{fig:V}
}
\end{figure}

\section{Thermodynamics}
Now that $f(z)$ is explicitly known in \nr{fz}, the standard formula
$4\pi T=-f'(z_h)$ with $dz/dQ$ from \nr{dzdQ} gives the horizon temperature
(when $Q_h$ is a dummy variable we leave out the index $h$).
Thus
\be
{1\over4\pi T}=b^3\int_Q^\infty dQ{1\over b^4W}
\ee
or
\ba
{\pi T(Q)\over\Lambda}&=&{\exp(-3Q)\over I(4Q,a)}\la{teequu}\\
&=&\exp\left(Q+{a\over4Q}\right)\left[1-{a\over16Q^2}+{a\over32Q^3}-
{3a\over 128Q^4}+{12a+a^2\over512Q^5}+..\right].
\la{tasymp}
\ea
Conversely, Eq. \nr{teequu} gives $Q_h=Q_h(\pi T/\Lambda)$.
The large-$T$ asymptotic expansion of this relation is, inverting \nr{tasymp}
and denoting $L\equiv \log(\pi T/\Lambda)$,
\be
Q_h=L-{a\over4L}+{a\over16L^2}-{a+2a^2\over32L^3}+{12a+25a^2\over512L^4}+..\,.
\la{QhL}
\ee

A particular role is also played by $dT/dQ$. In fact, below in \nr{cs} we shall
prove that $dT/dQ=3c_s^2(T) T$.

The relation $T=T(Q)$ is plotted in Fig. \ref{fig:TQ} in the small $Q$ region.
At large $Q$ the asymptotic expansion \nr{tasymp} is very accurate and
$T\sim\exp(Q)$. At $Q=0$
\be
{\pi T(0)\over\Lambda}= {1\over2\sqrt{a}K_1(2\sqrt{a})},
\quad {\pi\over\Lambda}\,{dT(0)\over dQ}=-3 {\pi\over\Lambda}T(0).\la{T0}
\ee
Thus $dT/dQ$ has to change sign at some $Q=Q_\rmi{min}$. At this point
there is a minimum of $T(Q)$ and $c_s^2=0$. In the more
general analysis of \cite{kiri3} the branch with $dT/dQ<0$ is called the
small black hole branch, the one with $dT/dQ>0$ is the big black hole one.
In Section \ref{1storder} we shall discuss the case when 
$T$ can grow to infinity in the small BH branch, now it is
bounded by \nr{T0}.

From the form \nr{ansatz} of the metric it
follows that the black hole entropy density is
\be
s={S\over V_3}={1\over4G_5}b^3(Q_h)={b_0^3\over 4G_5} \exp(3Q_h)
={\CL^3\over 4G_5}\Lambda^3\exp(3Q_h(T)).
\ee
Scaling with $T$, from \nr{teequu}, \nr{tasymp} and \nr{QhL},
\ba
{s\over T^3}&=&{(\pi\CL)^3\over4G_5}[\exp(4Q_h)I(4Q_h,a)]^3\nn
&=&
{(\pi\CL)^3\over4G_5}\exp\left({-3a\over4Q_h}\right)\,
\left(1+{3a\over16Q_h^2}-{3a\over32Q_h^3}+
{3a^2+9a\over 256Q_h^4}- {15a^2+36a\over512Q_h^5}+..\right)\\
&=&{(\pi\CL)^3\over4G_5}\left(1-{3a\over4L}+{9a^2+6a\over32L^2}-
{9a^3+24a^2+12a\over128L^3}+..\right),\qquad L\equiv\log{\pi T\over\Lambda}.
\la{entropy}
\ea

\begin{figure}[!tb]
\begin{center}

\includegraphics[width=0.45\textwidth]{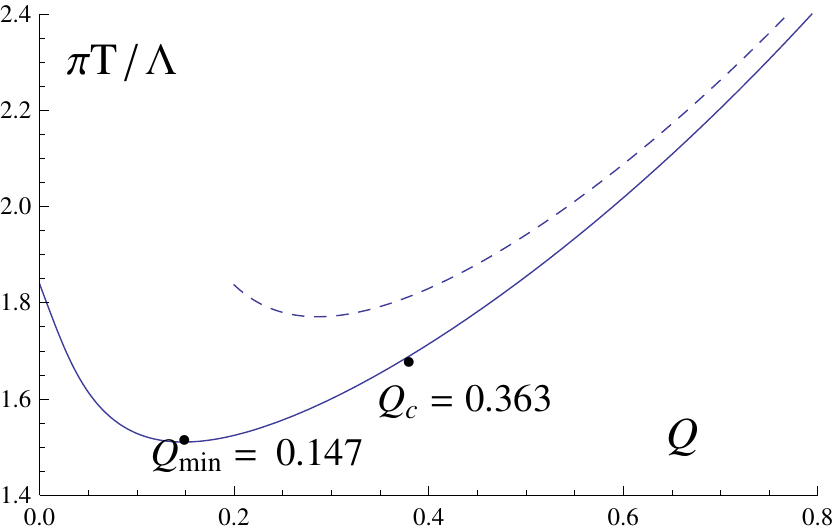}
\includegraphics[width=0.45\textwidth]{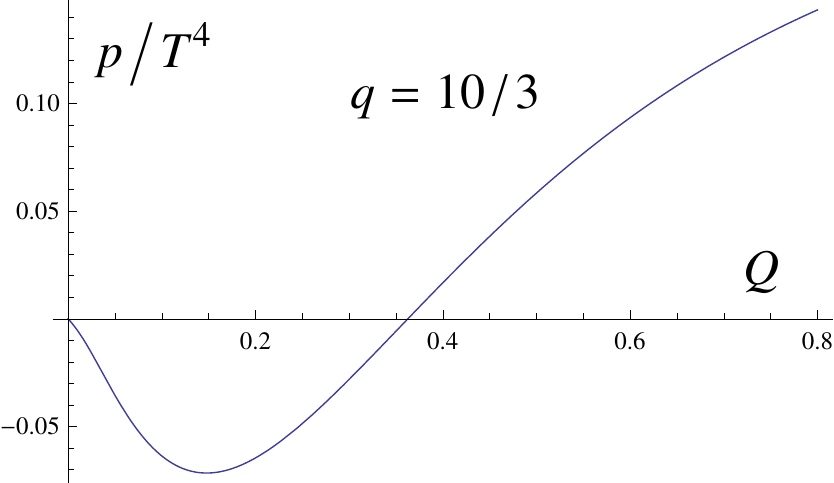}
\end{center}

\caption{\small $T(Q)$ from \nr{teequu} (left panel) and $p(T)/T^4$ from \nr{pee}
(right panel) with numbers corresponding to $q=10/3,\,a=16/49$. The pressure vanishes
at $p(Q_c=0.363)=0$, the corresponding value of $T$ is $\pi T_c/\Lambda=1.67$.
At this point $c_s^2=0.707/3$. The dashed line is the leading large $Q$
approximation $\exp[Q+a/(4Q)]$.
\la{fig:TQ}
}
\end{figure}

The full bulk thermodynamics is known when the (minus) free energy density or
pressure $p(T)$ is known. The rest then follows from relations like
\be
s(T)=p'(T),\quad\epsilon(T)=Ts-p,\quad {\epsilon-3p\over T^4}=
T{\partial\over\partial T}{p\over T^4},\quad c_s^2={dp\over d\epsilon}
={s\over Ts'(T)}.
\ee

Now that $s(T)$ is known, $p(T)$ can be obtained by integrating $s(T)=p'(T)$. The
key observation in \cite{kiri3,kiri4} now is that the correct constant of
integration is obtained by integrating over the entire curve $T=T(Q)$
starting the integration at the value
$T(0)$ corresponding to $Q=0$ in Fig. \ref{fig:TQ}. Thus one has
\be
p(T)=\int^T dT s(T)={b_0^3\over4G_5}\int_0^{Q(T)}dQ\,{dT\over dQ}\exp(3Q),
\qquad b_0=\CL\Lambda.
\la{pee}
\ee

Since $dT/dQ$ starts negative, $p(T)$ starts
decreasing (see Fig. \ref{fig:TQ}), but soon starts
growing and crosses the value zero at some $T$ which defines the transition
temperature between the phase we are discussing and the phase corresponding
to $f=1$ in the metric ansatz \nr{ansatz}. Since this low $T$ phase has $p=0$,
the transition temperature is determined by
\be
p(Q_c)=p(Q(T_c))=0,\la{tiisii}
\ee
where the value of $T_c={\rm number}\times \Lambda$ is obtained from \nr{teequu}.

From \nr{pee} one further computes
\ba
\epsilon(T)={b_0^3\over4G_5}\left[T\exp(3Q)-p\right]&=&
{b_0^3\over4G_5}\left[3\int_0^{Q(T)}dQ\,\,T(Q)\,\exp(3Q)+T(0)\right],\la{epsil}\\
\epsilon(T)-3p(T)&=&
{b_0^3\over4G_5}\left[3\int_0^{Q(T)}dQ\,\left(T-{dT\over dQ}\right)\,\exp(3Q)+T(0)\right]
\ea
and for the sound velocity
\be
c_s^2={1\over3T}\,{dT\over dQ}=\fr13\left[4{\pi T\over\Lambda}
\exp\left(-Q-{a\over 4Q}\right)-3\right]. \la{cs}
\ee

The large $Q$ or large $L=\log(\pi T/\Lambda)$ expansions are
\ba
{p\over T^4}&=& {(\pi\CL)^3\over4G_5}\left(\fr14-{3a\over16Q}+{9a^2\over128Q^2}
-{9a^3+12a\over512Q^3}+{27a^4+96a^2\over8192Q^4}+..\right)\nn
&=&{(\pi\CL)^3\over4G_5}\left(\fr14-{3a\over16L}+{9a^2\over128L^2}
-{9a^3+24a^2+12a\over512L^3}+..\right),\la{pT4NL}\\
{\epsilon\over T^4}&=&3{(\pi\CL)^3\over4G_5}\left(\fr14-{3a\over16Q}+{9a^2+8a\over128Q^2}
-{9a^3+24a^2+12a\over512Q^3}+..\right)\nn
&=&3{(\pi\CL)^3\over4G_5}\left(\fr14-{3a\over16L}+{9a^2+8a\over128L^2}
-{9a^3+48a^2+12a\over512L^3}+..\right),\\
{\epsilon-3p\over T^4}&=&3{(\pi\CL)^3\over4G_5}\left({a\over16Q^2}-{3a^2\over64Q^3}+
{9a^3+8a^2+12a\over 512Q^3}+..\right)\nn
&=&3{(\pi\CL)^3\over4G_5}\left({a\over16L^2}-{3a^2\over64L^3}+
{9a^3+24a^2+12a\over 512L^4}+..\right),\la{largeTim}\\
c_s^2&=&{s\over Ts'(T)}=\fr13\left(1-{a\over 4Q^2}+{a\over 8Q^3}+... \right).
\ea
Note that to plot the asymptotic expansions vs $T/T_c$ one must write
\be
\log{\pi T\over\Lambda}=\log\left({T\over T_c}{\pi T_c\over\Lambda}\right)
\ee
and determine $\pi T_c/\Lambda$ from \nr{tiisii}.

\begin{figure}[!tb]
\begin{center}

\includegraphics[width=0.46\textwidth]{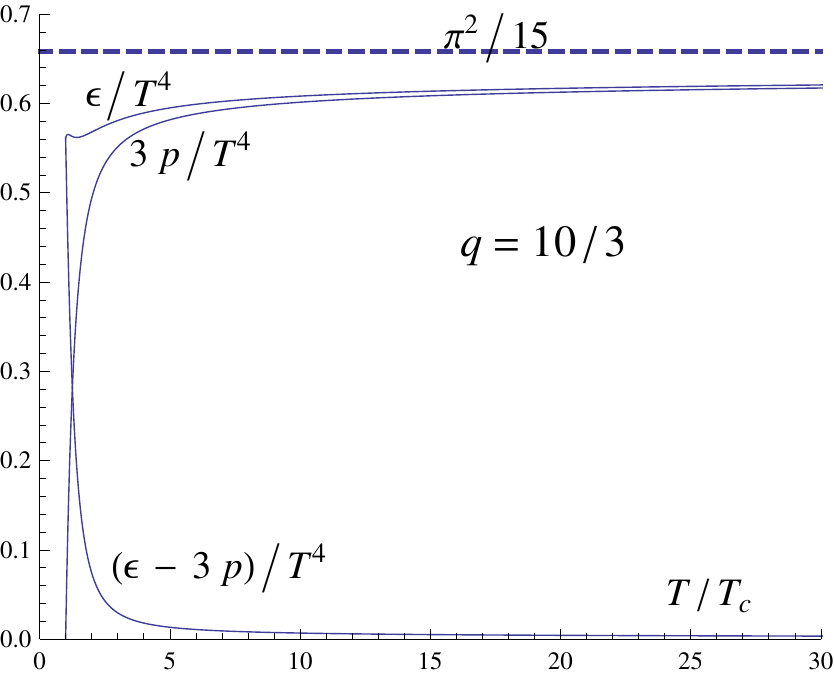}\hfill
\includegraphics[width=0.46\textwidth]{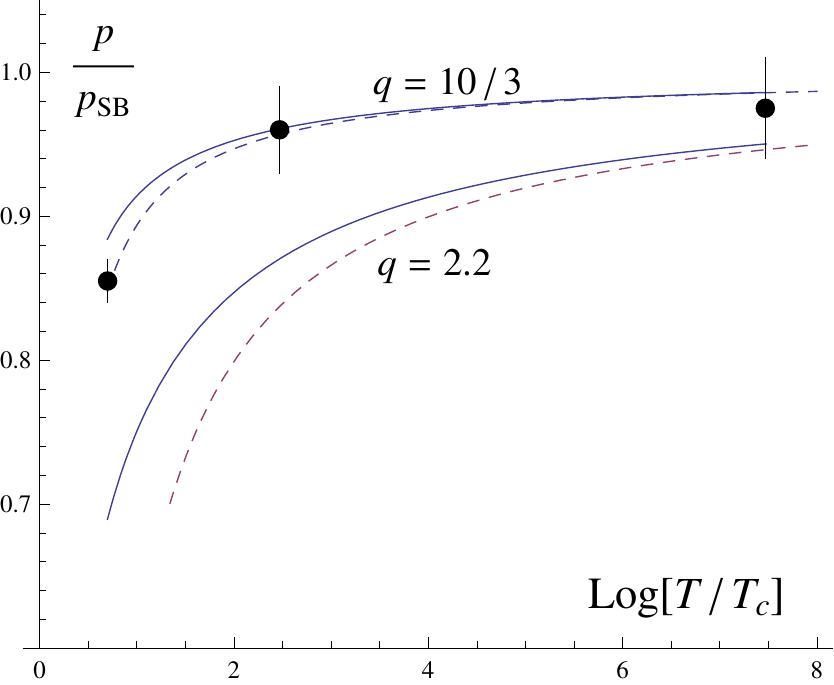}
\end{center}

\caption{\small Left panel: $\epsilon/T^4$, $3p/T^4$ and the scaled interaction measure
up to $30T_c$ for $q=10/3$. Here and in other
bulk thermo figures quantity/$N_c^2$ is plotted. The dashed line shows
the limit at $T\to\infty$. Right panel: $p(T)$ at very large $T$ scaled
by the $T\to\infty$ limit.
The two data points at $T=300T_c$ and $T=3\cdot 10^7T_c$ are from
\cite{fodor}, the third is the highest from \cite{boyd}.
The dashed lines are the asymptotic expansions \nr{pT4NL} including
only the $-3a/4\log(T)$ term.
}\la{fig4}
\end{figure}

\begin{figure}[!tb]
\begin{center}

\includegraphics[width=0.48\textwidth]{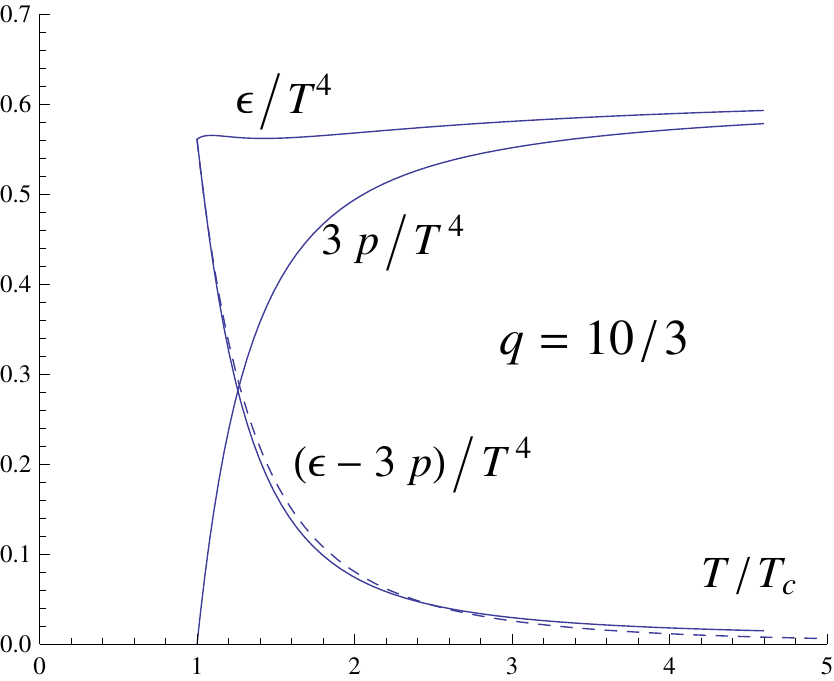}\hfill
\includegraphics[width=0.48\textwidth]{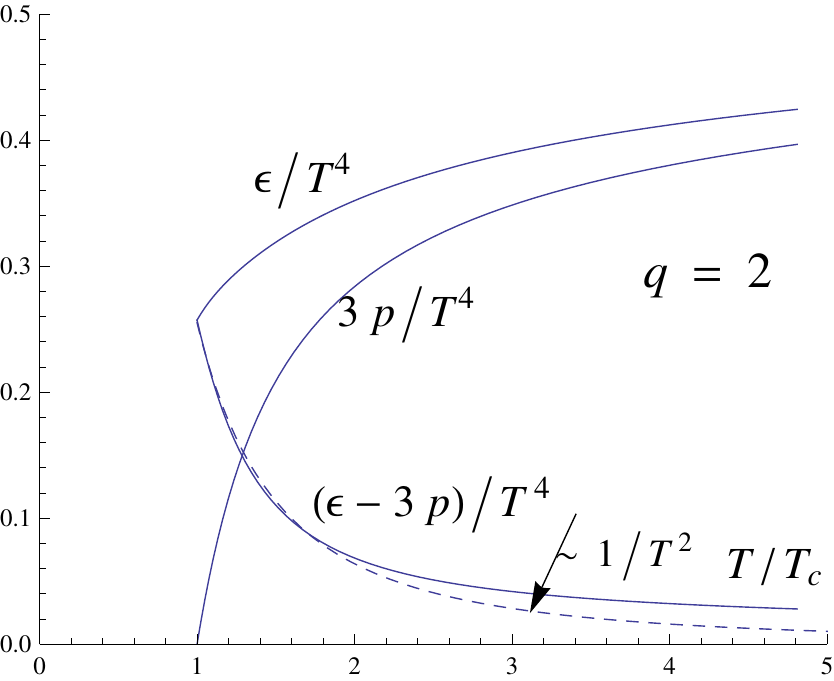}
\end{center}

\caption{\small $\epsilon/T^4$, $3p/T^4$ and the interaction measure up to $5T_c$ for
$q=10/3$ (left) and $q=2$ (right). For $q=10/3$ the
interaction measure decreases $\sim 1/T^{2.8}$ above $T_c$ (dashed line), for $q=2$
it decreases $\sim 1/T^2$ for $T_c<T\lsim 2T_c$ (dashed line), in agreement with data,
but more slowly for $T\gsim2T_c$ (see also Fig.~\ref{fig8}).
}\la{fig5}
\end{figure}

\begin{figure}[!tb]
\begin{center}

\includegraphics[width=0.45\textwidth]{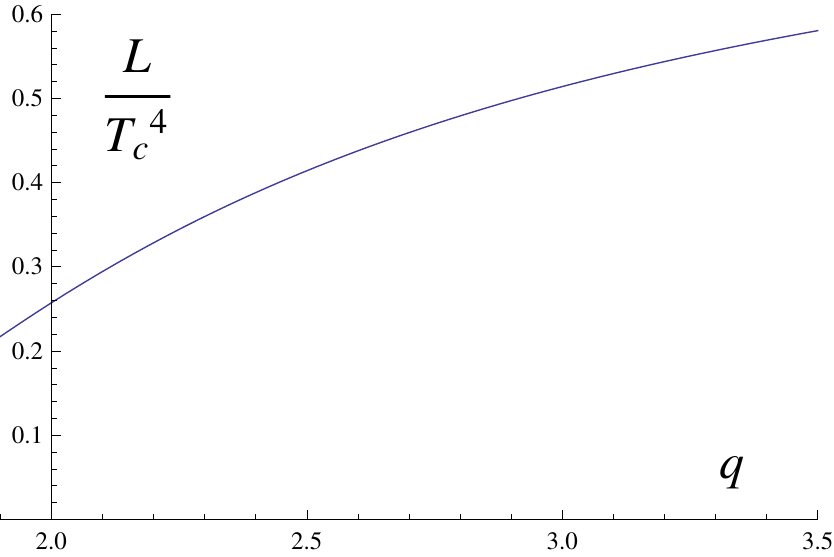}\hfill
\includegraphics[width=0.45\textwidth]{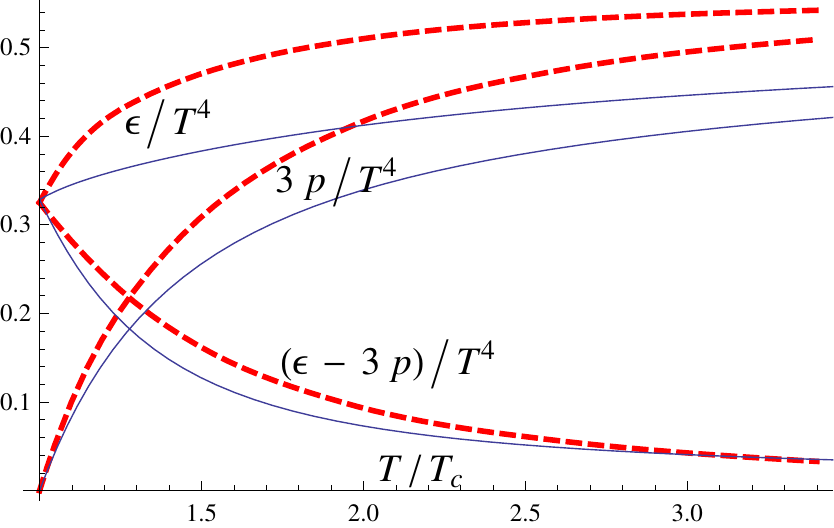}
\end{center}

\caption{\small Left panel: The latent heat plotted as a function of $q$. $L$ vanishes
at $q=1$. The lattice
value at $N_c\to\infty$ is 0.34(6) \cite{teper1} and is obtained for $q=2.2$. Right panel:
Lattice data for $N_c\gg1$ (dashed lines) \cite{panero} compared with model prediction for $q=2.2$.
}\la{fig7}
\end{figure}

\begin{figure}[!tb]
\begin{center}

\includegraphics[width=0.45\textwidth]{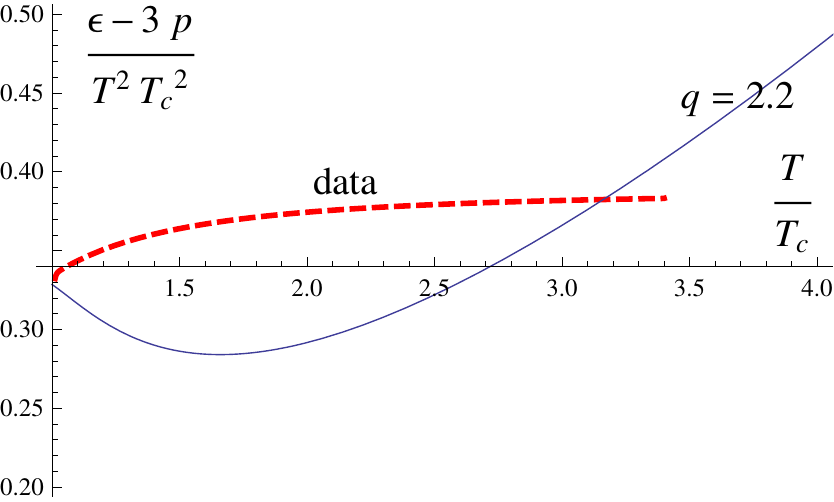}\hfill
\includegraphics[width=0.45\textwidth]{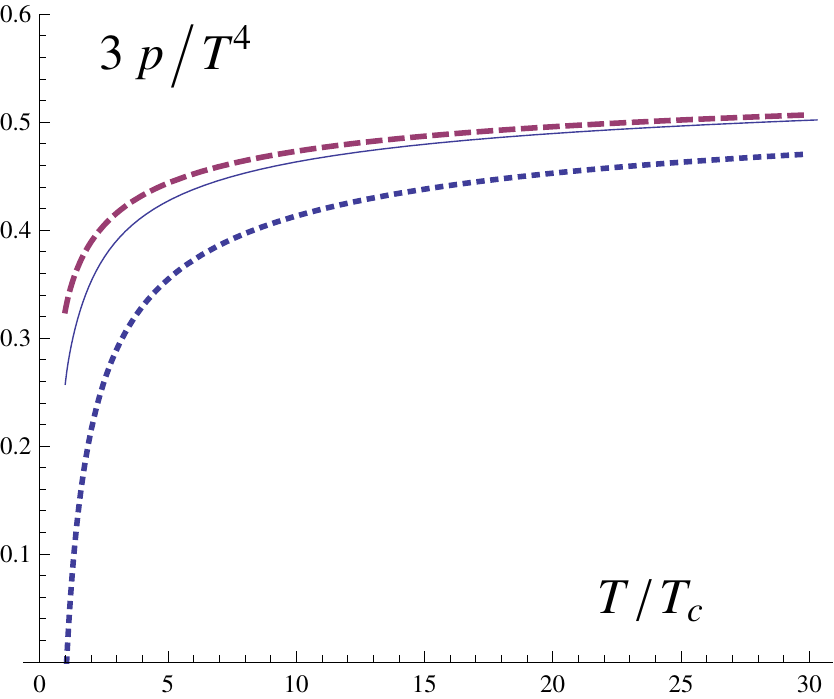}
\end{center}

\caption{\small Left panel: Data and model with $q=2.2$ for the interaction measure scaled by
$T^2$. Right panel:
$3p/T^4$ for $q=2$ plotted vs. $T/T_c$. Continuous curve (exact), dotted curve (Eq.
\nr{pT4NL} up to 1/log), dashed curve (Eq. \nr{pT4NL} up to 1/log$^3$).
}\la{fig8}
\end{figure}

\section{Comparison with lattice data for hot SU($N_c$) gauge theory}
Our model is so far based on a very simple powerlike in $\lambda$ (exponential in $\phi$)
ansatz for the beta function, from which the scalar potential is derived,
and one cannot expect detailed agreement with data. Nevertheless, we find that
on a qualitative level the main features of the data are reproduced to
surprising extent.


To analyse the bulk quantities one has to fix the value of the parameter
$\CL^3/G_5$. We fix it to the pressure of ideal gluon gas,
\be
p_\rmi{SB}=2(N_c^2-1){\pi^2\over90}T^4.
\la{SB}
\ee
Comparing with \nr{pT4NL} one has
\be
{(\pi\CL)^3\over4G_5}={4\pi^2\over45}N_c^2.
\la{normalis}
\ee
This is less by a factor $8/45$ relative to the value for ${\cal N}=4$ SYM,
$2/15$ comes from the reduction of the number of degrees of freedom and $4/3$
from going from strongly interacting to ideal systems.
Bulk thermodynamic quantities are very accurately proportional to $N_c^2$
\cite{teper1,teper2,panero} so that we shall simply divide by this factor.

Using this normalisation Figs. \ref{fig4} and \ref{fig5} show $\epsilon/T^4,\,p/T^4,
(\epsilon-3p)/T^4$ as functions of $T/T_c$
evaluated numerically from the above equations for $q=10/3$ and $q=2$. Fig.~\ref{fig4}
also shows $p(T)$ at very large $T$ scaled by the $T\to\infty$ limit \nr{SB}.

At $T_c$, $p=p(T_c)=0$ but $\epsilon(T_c)$ has the nonzero value given by \nr{epsil}.
Since $p=\epsilon=0$ for the low $T$ phase, $p$ is continuous, by construction,
but $\epsilon$ jumps. This gives the latent heat
\be
{L\over T_c^4}\equiv {\epsilon(T_c)\over T_c^4}={s(T_c)\over T_c^3}=
{4\pi^2\over45}N_c^2\left[\exp(4Q_c) I(4Q_c, a)\right]^3.
\la{latent}
\ee
This is plotted in Fig. \ref{fig7} as a function of $q$. The latent heat vanishes for
$q=1$, the transition goes over to a second order one.
The lattice value \cite{teper1}
$L/T_c^4=0.34(6)N_c^2$ would be obtained for $q\approx2.2$. Choosing this value of $q$,
Fig. \ref{fig7} also compares the $T$ dependence of the bulk quantities with lattice data
extrapolated to $N_c\to\infty$ \cite{panero}.

One well known characteristic of $(\epsilon-3p)/T^4$ is that it
decreases $\sim 1/T^2$ for $T\gsim T_c$. This is checked in Fig. \ref{fig8} for the
data and for the model with $q=2.2$.
As seen in Fig.~\ref{fig5}, for $q=10/3$ the model result decreases faster,
$\sim 1/T^{2.8}$. Continuing to still larger $T$,
according to \nr{largeTim} the leading large $T$ behaviour of the interaction measure is
\be
{\epsilon-3p\over T^4}={4\pi^2N_c^2\over135(q-1)^2}{1\over\log^2(\pi T/\Lambda)},
\la{intmeasNL}
\ee
a slight generalisation of Eq. (I.3) in \cite{kiri3}.

The asymptotic expansion \nr{pT4NL} looks much like a perturbative expansion
in $g^2$ and it is
of interest to compare the leading corrections quantitatively - to higher orders
the finite $T$ perturbative expansion proceeds in powers of $g$.
Consider first the interaction measure. In perturbation theory,
to leading order, by taking $T\partial/\partial T$ of Eq. (6.1) of \cite{g6g},
\be
{\epsilon-3p\over T^4}=-{(N_c^2-1)N_c\over144}T{dg^2(T)\over dT}
={\pi^2(N_c^2-1)\over 66}{1\over\log^2(\pi T/\Lambda_\tinymsbar)}.
\ee
This is very much similar to \nr{intmeasNL}, in fact, coincides for
$q=\sqrt{88/45}+1\approx2.4$. Consider then the pressure, for which
the two expansions are
\be
{p\over p_\rmi{SB}}=1-{5g^2N_c\over 16\pi^2}+{80\over\sqrt3}
\left({g^2N_c\over 16\pi^2}\right)^{3/2}+..=1-{4\over3(q-1)^2\log(\pi T/\Lambda)}+..,
\ee
where also the $g^3$ plasmon term is included in the perturbative expansion.
Comparing the leading terms with 1-loop running for $g^2(T)$,
the two expansions again coincide for
$q=\sqrt{88/45}+1\approx2.4$. However, the plasmon term is quantitatively
very important, smaller than the $g^2$ term only for $T>10^5T_c$, which
explains why the $q=10/3$ term agrees with the perturbative result, see Fig.~\ref{fig4}.
This figure also shows to what extent the leading order term of the expansion \nr{pT4NL}
agrees with the numerical computation.
How still further terms improve the agreement is seen from
Fig. \ref{fig8}.

From the above one sees that $q=10/3$ gives a good fit to SU($N_c$) thermodynamics for
$T\gsim10 T_c$; for smaller $T$ the curves are too high and the latent heat too big.
This indicates that one has to modify the ansatz in the IR large $\lambda$ region,
the excellent fit in \cite{kiri4} contained several parameters. In the next two sections
we present systematic generalisations of \nr{betaappro}, which will be seen to lead first to
vanishing latent heat, a continuous transition and then again to a first order transition.

\section{Continuous or second order transition}\la{conttrans}
As discussed above, the transition approaches a continuous one when $q\to1$. To study this
case it is more realistic not to take $q\to1$ in \nr{betaappro} but to take a
beta function which only approaches the marginally 
confining $q=1$ form $-\fra32\lambda$ at large
$\lambda$:
\be
\beta(\lambda)={-\beta_0\lambda^q\over 1+\fra23\beta_0\lambda^{q-1}}.
\la{betaappro2}
\ee
Even for this case the solution of the gravity equations \nr{eq1}-\nr{crucial} is
surprisingly simple. Denoting again (but now this is not $\log(b/b_0)$)
\be
Q={1\over(q-1)\beta_0\lambda^{q-1}}
\la{betalam2}
\ee
and recalling the abbreviations
\be
A=\fr12\sqrt{a}={2\over 3(q-1)}
\ee
we have
\ba
b&=&b_0\, e^Q\left({Q\over Q_0}\right)^A,\la{b2}\\
W&=&{1\over\CL}\left(1+{A\over Q}\right)^A\la{W2},
\ea
where $Q_0$ is the scale at which $b$ has the value $b_0 e^{Q_0}$. In
Section \ref{1storder} we shall choose $Q_0=A$. 
Note that now $b$ vanishes in the IR proportionally to a power of $Q$
or of $1/\lambda$.

Using $db/dz=-b^2W$ from \nr{dzdQ} and $4\pi T=-\dot f(z_h)$ from \nr{fz} one
finds the temperature
\be
{\Lambda\over \pi T}\,{1\over 4Q_0^A}=
e^{3Q}\,Q^{3A}\int_{Q}^\infty dy{y^{-3A-1}\over (y+A)^{A-1}}\,e^{-4 y}
\equiv{1\over \hat T(Q,A)},
\la{tee2}
\ee
the pressure ($Q_0$ cancels from here)
\be
{p\over T^4}={(4\pi\CL)^3\over 4G_5}\,{1\over \hat T^4}\int_0^Q dQ {d\hat T\over dQ}\,
e^{3Q}\,Q^{3A}
\la{p2ndo}
\ee
and sound velocity
\be
c_s^2={1\over3}{Q\over A+Q}{dT\over TdQ}.
\ee
A numerical example of the scaled bulk quantities for $q=10/3$ is shown in Fig.~\ref{fig9}.

Now that the transition is of second order the minimum of $T(Q)$ is at
$Q=0$, beyond that $dT/dQ>0$. $T(Q=0)$ is calculable and defines the critical temperature
\be
{\pi T_c\over\Lambda}={3\over4}\left({A\over Q_0}\right)^A.
\ee
To find the behavior of $T(Q)$ near $Q=0$ or 
near $T_c$ one must do partial integrations with $dy^{-3A}$
to increase the power of $y$ in
\nr{tee2}. For $0<3A<1,\,q>3$ one partial integration is enough to make the integral
converge for $Q\to0$. Expanding the remainder in
$Q$ one obtains
\ba
&&\hat T(Q,A)/\hat T(0,A)-1=T/T_c-1\equiv t
\nn
&& =
A^{A-1}\int_0^\infty dy\, y^{-3A}\,e^{-4y}{5A-1+4y\over(A+y)^A}\,\,Q^{3A}
-{6A\over 1-3A}Q+{\cal O}(Q^2),
\la{critreg}
\ea
where the integral is positive for $0<3A<1$. For this range of $3A$ the first
term is dominant in the critical region $t\to0$ so that
\be
t\sim Q^{3A},\qquad Q\sim t^{1/3A}\qquad 0<3A<1.
\ee
For $1<3A<2,\,2<q<3$ two partial integrations are needed and
\be
t = {6A\over 3A-1}\,Q - {\rm Integral}\,Q^{3A}+{\cal O}(Q^2),
\ee
where the Integral, similar to the one in \nr{critreg}, is positive. Thus
now and for all larger values of $A$ and smaller values of $q$:
\be
t\sim Q,\qquad 1<3A<2,\quad 2<q<3.
\ee
When converted to the critical behavior of the free energy ($=-p$) one
has
\be
f\sim t^2,\quad 3A<1\qquad f\sim t^{3A+1},\quad 3A>1
\ee
so that the critical exponent in $C_V\sim f''(t)\sim t^{-\alpha}$ is
$\alpha=0\,\,(=1-3A)$ for $3A<1,\,\,(3A>1)$. Since $f''(t)$ is not divergent,
the transition is a continuous one. For the SU(2) finite $T$ transition, of
3d Ising universality class, one has $\alpha=0.12$ \cite{engels}, a
mild divergence. A continuous transition appears also if one takes
SU(3) gauge theory with effectively 3 flavours of infinite mass fermions
and reduces the mass of one fermion, i.e., studies one-flavor QCD.
For sufficiently small mass the
transition becomes a continuous one \cite{alexandrou}, the end point again
being of Ising universality class.

With the modified beta function \nr{betaappro2} with $q=10/3$ we have now
reduced the latent heat to zero. The work in \cite{kiri3,kiri4}
indicates what the modification should be if one wants a first order transition
but still maintains the leading asymptotic behavior $-\fra32\lambda$: the asymptotic behavior
at large $\lambda$ should be
\be
\beta\to-\fra32\lambda\left(1+{\alpha\over\log\lambda}\right)
\ee
with\footnote{In the notation of \cite{kiri3}, this parameter was denoted by
$\fra34\alpha/(\alpha-1)$, $\alpha\ge1$} 
$\alpha>0$. We shall in the next section derive analytic formulas even for this
case. In principle, one could then redo
the critical analysis above so that the parameter $\alpha$ would effectively be an
external field which could drive the transition to a first order one
and more critical indices would appear.

\begin{figure}[!tb]
\begin{center}

\includegraphics[width=0.6\textwidth]{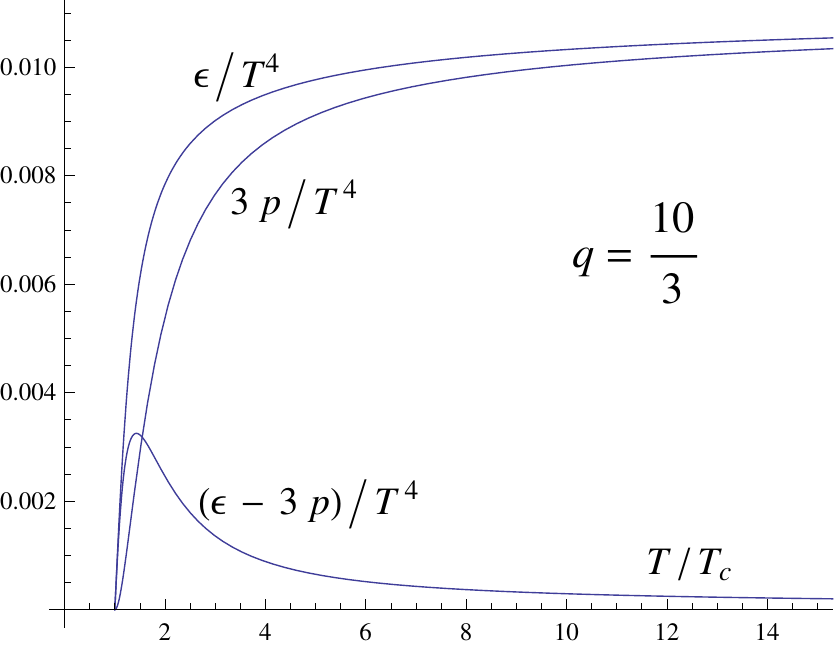}
\end{center}

\caption{\small Scaled bulk thermodynamic quantities for the beta function \nr{betalam2},
leading to a continuous transition, 
with $q=10/3$. If the curves were normalised as in \nr{normalis}, they should be multiplied
by $256\pi^2/15\approx 168$.
}\la{fig9}
\end{figure}

\section{First order transition again}\la{1storder}
Consider now the following ansatz:
\ba
\beta(\lambda)&=&{-\beta_0\lambda^q\over 1+\fra23\beta_0\lambda^{q-1}}
\left[1+ \alpha(q-1){\log(1+\fra23\beta_0\lambda^{q-1})\over
\log^2(1+\fra23\beta_0\lambda^{q-1})+1}\right]\\
&=&{-\lambda\over q-1}\,{1\over Q+A}\left[1+\alpha(q-1){\log(1+A/Q)\over \log^2(1+A/Q)+1}\right],
\la{betaappro3}
\ea
where $Q,\,A$ are defined as in the previous section.
The modification factor relative to \nr{betaappro2} is so constructed that in the IR,
$\lambda\to\infty$, it is $1+\alpha/\log\lambda+...$,
in the UV it is $1+{\cal O}(\lambda^{q-1})$ and also so that
$W$ is integrable in closed form from \nr{W}:
\be
W={1\over\CL}\left(1+{A\over Q}\right)^A\,
\left[\log^2\left(1+{A\over Q}\right)+1\right]^{\fra13\alpha},
\ee
The expression for $b$ from \nr{crucial},
\be
\log{b\over b_0}=\int_{\lambda_0}^\lambda{d\lambda\over\beta}=
\int_A^Q{dQ\over Q}{-\lambda\over(q-1)\beta(\lambda)}=
\int_{A}^Q dy\left(1+{A\over y}\right){1\over 1+
{\alpha(q-1)\log(1+A/y)\over \log^2(1+A/y)+1}},
\la{betalamfull}
\ee
is also surprisingly simple, though 
not integrable in closed form. The lower limit of the $Q$ integral has here been fixed
to be $A$, i.e., $b(Q=A)=b_0$. In the IR, $Q\to0$, this implies
\be
{b\over b_0}=\left({Q\over A}\right)^A\,\left({1\over\alpha(q-1)}
\log\fr{A}{Q}+1\right)^{\fra23\alpha}.
\la{b3}
\ee
This is the key modification relative to \nr{b2}: $b$ vanishes in the IR
but more slowly by a power of $\log Q$ than an ansatz with $\beta\to-\fra32\lambda$.

The modified IR behavior has several consequences. Consider first the $z$ dependence
of the field configurations. For this one has to integrate $z=z(Q)$ from
$dz=-db/(b^2W)$ so that, as a generalisation of \nr{zQ},
\be
\Lambda z = \int_{Q(z)}^\infty dQ\,{d\log b\over dQ}\,{1\over bW}.
\ee
Studying the limit $Q(z)\to0$ one sees that this integral diverges if $\fra43\alpha<1$:
\be
\Lambda z\sim \log(A/Q)^{1-\fra43\alpha},
\ee
otherwise it is finite. Thus, if $\alpha<\fra34$ the range of $z$ is unbounded.

Similarly, for the temperature we have
\be
{1\over4\pi T}=b^3\int_0^{z_h}{dz\over b^3}=b^3\int_{Q(T)}^\infty dQ\,{d\log b\over dQ}
{1\over b^4W}
\la{Tgeneral}
\ee
and counting IR logs one finds that the power of $\log(1/Q)$ is $2\alpha$
from the outside $b^3$, $-8\alpha/3-2\alpha/3$ from the inside $1/(b^4W)$
so that in the IR $T\sim\log^{4\alpha/3}(1/Q)$. Thus $T$ diverges logarithmically
when $Q\to0$;
with the beta function \nr{betaappro} $T(Q=0)$ was finite but there was a minimum,
Fig.~\ref{fig:TQ}. Thus there again is a minimum of $T(Q)$ and a branch with $dT/dQ<0$, 
so that $p(T)$ behaves as in Fig.~\ref{fig:TQ} and the transition becomes of first
order. Full thermodynamics
in terms of two parameters, $q>1$ and $\alpha>0$, is now given by the formulas,
generalisations of \nr{pee} and \nr{epsil}:
\ba
s(T)&=&{1\over 4G_5}b^3(Q(T)),\\
p(T)&=&\int^T dT s(T)={1\over4G_5}\int_0^{Q(T)}dQ\,{dT\over dQ}b^3(Q),\la{pee2}\\
\epsilon(T)&=&
{3\over4G_5}\int_0^{Q(T)}dQ\,\,T(Q)\,b^3(Q){d\log b\over dQ},\la{epsil2}\\
\epsilon(T)-3p(T)&=&
{3\over4G_5}\int_0^{Q(T)}dQ\,\left(T{d\log b\over dQ}-{dT\over dQ}\right)\,b^3(Q),\\
c_s^2&=&{1\over3T}\,{dT\over dQ}{dQ\over d\log b}=\fr13\left(4{\pi T\over bW}-3\right). \la{cs2}
\ea
Note that $T(0)$ in \nr{epsil} is actually $T(0)b^3(0)$ which now vanishes. The procedure
again is that one first finds at what value of $Q=Q_c$ the pressure vanishes: $p(Q_c)=0$.
From \nr{Tgeneral} one next finds the corresponding value of $T=T_c=T(Q_c)$. Finally one
plots parametrically with $Q>Q_c$ as the parameter the
required thermodynamic quantity as the $y$ axis and $T(Q)/T_c$ as $x$ axis.

\begin{figure}[!tb]
\begin{center}

\includegraphics[width=0.46\textwidth]{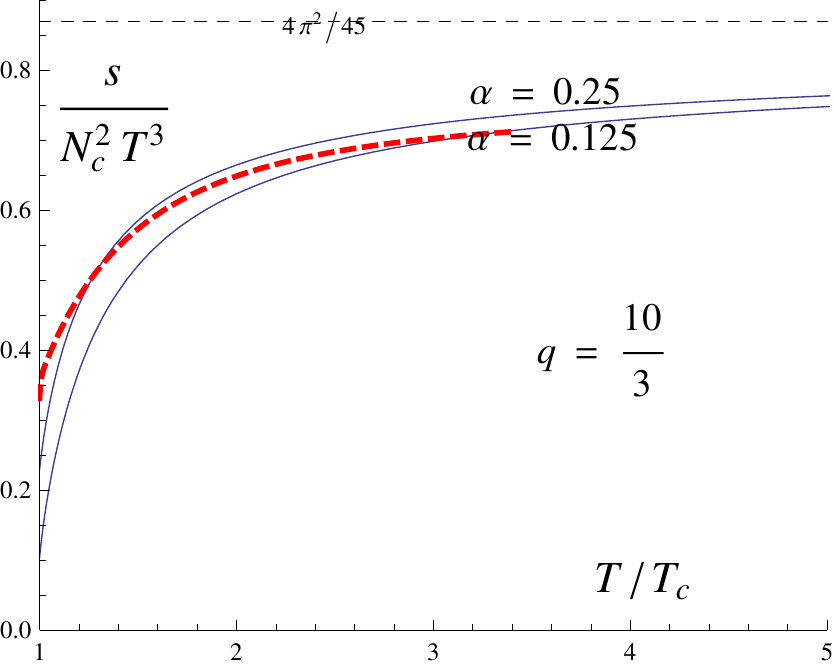}\hfill
\includegraphics[width=0.46\textwidth]{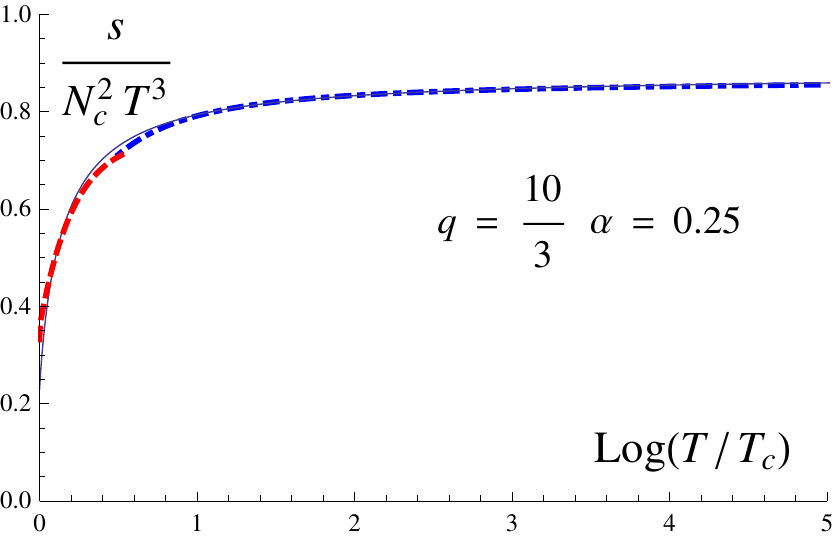}
\end{center}

\caption{\small Entropy density scaled by $T^3N_c^2$ 
(the asymptotic $T\to\infty$ value is $4\pi^2/45$)
for the beta function \nr{betaappro3}
with $q=10/3$ and $\alpha=0.25,\,\,0.125$ and for $T<5T_c$ (left panel)
or $T<10^5T_c$ (right panel).
The dashed curve gives the $N_c\to\infty$ limit of lattice
data \cite{panero} and the dotdashed one a QCD perturbative result (see text).
}\la{fig10}
\end{figure}

We refer a detailed comparison with lattice data in the few $T_c$ range
and with QCD perturbation theory at large $T$ to later work. As a preliminary
step, Fig.~\ref{fig10} compares the entropy density scaled by $N_c^2T^3$ 
obtained for $q=10/3$
with the $N_c\to\infty$ limit of lattice data (known \cite{panero} for
$T\le 3.4T_c$) and the QCD perturbative result
\footnote{The perturbative result is obtained by computing entropy density 
from the pressure in Eq.~(6.4) of \cite{g6g}, choosing the renormalisation
scale $\bar\mu=2\pi T$, the unknown 4-loop Linde coefficient $q_c=-3800$ and a
2-loop running coupling with $N_c=3$.} at large $T$. 
One sees that the too large latent heat observed in Fig.~\ref{fig5} for
this value of $q$ (which leads to good large $T$ behavior) is cured
by the modification of the IR region incorporated in the beta function
\nr{betaappro3}, if one chooses $\alpha$ to be in the range $0.1...0.2$. 
There is still some deviation very close to $T_c$, but overall the
agreement is excellent, as already noted in \cite{kiri4}. 

Finally, one may note that the latent heat vanishes when $\alpha\to0$
as a power of $\alpha$:
\be
{L\over N_c^2T_c^4} = 1.55\,\alpha^{1.30},\qquad q=\fra{10}3.
\ee
This is accurate up to $\alpha=0.1$.

\section{Conclusions}
In this paper, we have studied a simple gauge/gravity duality model for gauge theory
thermodynamics. Basically, the model is a modification of the thorough work carried
out in \cite{kiri1,kiri2,kiri3,kiri4}: instead of starting from a scalar field potential
$V(\phi)$ of the gravity side we start from the beta function on the gauge theory
side. Its chief virtue is simplicity, semianalytic formulas are obtained, no numerical
solutions of Einstein's equations are needed and the only 
essential parameters are the power $q>1$
of $\lambda=e^\phi$ in the beta function and the parameter $\alpha>0$
associated with the approach to the infrared limit. Notable is also that
the normalisation of $\lambda$ never enters, everything is expressed using
the combination $Q=1/((q-1)\beta_0\lambda^{q-1})$.

The key element in this analysis was that the three equations determining the three functions
$b(z),\,\phi(z),\,\lambda(z)$ were taken to be the two Einstein equations not containing
the potential $V(\phi)$ and the equation relating the beta function to $b(z),\,\lambda(z)$.
The Einstein equation containing the potential then serves simply to determine the
potential.

It is perhaps surprising that such a simple beta function as $\beta=-\beta_0\lambda^q$
reproduces main features of a first order transition so well. 
Detailed agreement with lattice data and QCD perturbation theory requires beta functions
constrained to approach $-\fra32\lambda(1+\alpha/\log\lambda)$ at large $\lambda$. 
Here $\alpha>0$ leads to a first order transition. 
Even for this case analytic expressions could be given.
They become particularly simple for the case $\alpha=0$, corresponding
to a continuous transition.

There are many directions into which one could develop this work. In particular, a
study of beta functions of the type $-\beta_0\lambda^q(1-\lambda/\lambda_*) $ containing
an infrared fixed point $\lambda_*$ suggests itself. Further, the model could be
applied to walking technicolor beta functions which only approach a fixed point
$\lambda_*$. Perhaps then one looses the virtue of simplicity.

\vspace{1cm}
{\it Acknowledgements}.
We thank E. Kiritsis and F. Nitti for explaining their model.

\end{document}